\documentclass[reprint,amsmath,amssymb,aps,pre,showpacs]{revtex4-1}
\usepackage{graphicx}
\usepackage{dcolumn}
\usepackage{bm}
\usepackage{hyperref}
\usepackage{bookmark}
\usepackage{subfigure}

\begin{document}
\title{Synchronization of nearly-identical dynamical systems. I Size instability}
\author{Suman Acharyya$^{a,}$}
\email{suman@prl.res.in}
\author{R. E. Amritkar$^{a,b,}$}
\email{amritkar@prl.res.in}
\affiliation{$^a$ Physical Research Laboratory, Ahmedabad, India \\
$^b$ Institute of Infrastructure, Technology, Research and Management, Ahmedabad, India}

\begin{abstract}
We study the generalized synchronization and its stability using master stability function (MSF), in a network of coupled nearly identical dynamical systems. We extend the MSF approach for the case of degenerate eigenvalues of the coupling matrix. Using the MSF we study the size-instability in star and ring networks for coupled nearly identical dynamical systems. In the star network of coupled R\"ossler systems we show that the critical size beyond which synchronization is unstable, can be increased by having a larger frequency for the central node of the star. For the ring network we show that the critical size is not significantly affected by parameter variations. The results are verified by explicit numerical calculations.
\end{abstract}


\maketitle

\section{\label{Introduction}Introduction}

In physical world networks are ubiquities. Many practical complex systems in natural and social sciences and also humanities can be modeled as networks of interacting systems. Recently, the study of such complex networks has attracted much attention~\cite{Barabasi2002,Dorogovtsev_book2003,Newman2003,Boccaletti2006}. The complex networks can be grouped into different types depending on their structure into some universal categories, such as random networks~\cite{Erdos-Renyi1960}, scale free networks~\cite{Barabasi1999}, small world networks~\cite{Watts1998}, etc.

When there are several interacting dynamical systems on networks, they can exhibit a rich variety of dynamical behavior such as synchronization~\cite{Kurths-book,Boccaletti2002,Arenas2008}, amplitude death~\cite{Reddy1998}, multistability~\cite{Kim1997}, chimera states~\cite{Abrams2004}, phase flip~\cite{Prasad2006}, etc. which a single dynamical system is unable to show. Occurrence of  synchronization between interacting dynamical units is an important and fundamental nonlinear phenomenon and the study of synchronization of coupled dynamical systems has attracted considerable attention in the past decades~\cite{Kurths-book,Boccaletti2002,Arenas2008,Fujisaka1983,Yamada1983,Afraimovich1986,Pecora1990,Heagy1994,Pecora1997,Rosenblum1996,Abarbanel1995,Abarbanel1996}. In particular, the nonlinear behavior of coupled chaotic systems tends to separate the nearby trajectories of the coupled systems, while a suitable coupling between them brings back the trajectories together. In this competition, when the later wins 
the coupled systems undergo synchronization. Synchronization of coupled dynamical systems can be
defined as a process where two or more coupled systems adjust their trajectories to a common behavior when they are coupled or driven by a common signal. In the context of coupled chaotic systems different types of synchronizations have been studied in the past years. These include complete or identical synchronization~\cite{Fujisaka1983,Yamada1983,Pecora1990}, phase synchronization~\cite{Rosenblum1996,Rosa1998}, lag synchronization~\cite{Rosenblum1997}, anticipatory synchronization~\cite{Ambika2009}, imperfect phase synchronization~\cite{Zaks1999}, generalized synchronization~\cite{Abarbanel1995,Abarbanel1996}, measure synchronization in Hamiltonian systems~\cite{Hampton1999} etc. Among these the simplest and the most studied is the complete synchronization which occurs in two or more coupled identical dynamical systems and is characterized by the equality of state variables of the interacting systems. The stability of synchronization is normally determined by the negativity of the largest transverse 
Lyapunov
exponent~\cite{Heagy1994,Pecora1998}. Pecora and Carroll have developed an elegant way, namely the master stability function (MSF) for analyzing the stability of complete synchronization for a network of identical dynamical systems~\cite{Pecora1998}. The MSF allows one to study the stability of synchronization of different networks using a single function and has been used widely for a comparative study of synchronization of different networks of identical dynamical systems~\cite{Fink2000,Barahona2002,Nishikawa2003,Zhao2005,Zhou2006,Huang2009,Zhu2010}.
It is shown that the small world network enhances synchronizability of a network of coupled identical systems~\cite{Barahona2002}. In Ref~\cite{Nishikawa2003} it is shown that having smaller network distance is not sufficient for performing best synchronizability properties, it is also required to have homogeneous degree distribution among the coupled dynamical systems. Ref.~\cite{Donetti2005} introduces a new family of graphs, namely the entangled networks, which show better synchronizability. These entangled networks are interwoven and have extremely homogeneous structures, i.e. the degree
distributions are very narrow.

The network of coupled identical dynamical systems is an ideal situation and in practical situations it is almost impossible to have a network of coupled identical dynamical systems. So it is important to study coupled nonidentical systems and see how their behavior compares with that of coupled identical systems. For the coupled nonidentical systems, one can not get complete synchronization.
Instead in this case the synchronization is of a generalized type where the state variables of the coupled dynamical systems are related with some functional relationship~\cite{Abarbanel1995,Abarbanel1996}. The nonidentical nature of the coupled systems can lead to desynchronization bursts which is known as the bubbling transition~\cite{AshwinPLA1994,RestrepoPRL2004}. After the desynchronization burst the system returns to the synchronized state. Sun et al.~\cite{SunEPL2009} have extended the master stability approach for nearly identical systems to calculate the deviation from the average trajectory and the deviation is shown to be bounded.

In this paper, we extend the MSF formalism to a system of nearly identical systems. By coupled nearly identical systems we mean systems which have a node dependent parameter (NDP). Preliminary results of this study were earlier reported in~\cite{Acharyya2012}. We then extend the MSF formalism to coupled nearly identical systems with degenerate eigenvalues of the coupling matrix. We also obtain MSF for systems with more than one NDP. We find that, in general, the stability of synchronization can be improved when we introduce non-identical nature in the coupled systems.

Next we use MSF to study size-instability of synchronization in well structured networks. By a well structured network we refer to a network in which the number of nodes can be changed without changing the basic structure and symmetry of the network, e.g. star network, ring network etc. The size-instability is the phenomena that there is a critical number of oscillators that can be coupled in a well structured network to obtain synchronization and beyond this critical number no stable synchronization can be seen. The phenomena of size-instability in identical oscillators is well known and has been studied widely~\cite{Heagy1994,TBohrPRL1989,MatiasPRL1998,RestrepoPRL2004,SenthilkumarPRE2010}. For the star network of coupled nearly identical systems, we find that it is possible to increase the critical number of nodes beyond which synchronization is unstable, by a judicious choice of NDP. In particular, for coupled R\"ossler systems the critical number can be increased by having a larger frequency for the 
central node. On the other hand, we find that for a ring network, the critical number of nodes is not significantly affected by an NDP. These results are verified by explicit numerical calculations. In part II of this study \cite{next}, we use MSF to construct optimized networks for better synchronization properties and study properties of these optimized networks.

\section{\label{stability}Stability of synchronization of coupled nearly identical systems}

For networks of coupled identical systems the stability of complete synchronization has been well analyzed. As discussed in the introduction, Pecora and Carroll (1998)~\cite{Pecora1998} introduced a master stability function (MSF) which can be calculated from master stability equations. Using the master stability function one can calculate the largest transverse Lyapunov exponent for a network and study and compare the stability properties of synchronization of different networks. For coupled nearly identical systems, the synchronization is of a generalized type. We now extend the MSF approach to coupled nearly identical systems.

\subsection{\label{MSF} Master Stability Function for nearly-identical systems}

In Ref.~\cite{Acharyya2012}, we have extended the formalism of MSF to coupled nearly-identical systems and in this subsection, we briefly review the same. This is done for the sake of completeness and also to establish the notation. We start by considering a network of $N$ coupled dynamical systems as
\begin{equation}
\dot{x}^i = f(x^i,r^i) + \varepsilon\sum_{j=1}^N g_{ij} h(x^j);\; i=1,...,N
\label{N-systems}
\end{equation}
where $x^i(\in R^m)$ is the $m$-dimensional state vector of system $i$, $r^i$ is the node dependent parameter (NDP) which makes the systems nonidentical, $f:R^m \rightarrow R^m$ and $h: R^m \rightarrow R^m$ give respectively the dynamical evolution of a single system and the coupling function, $G=[g_{ij}]$ is the coupling matrix and $\varepsilon$ is the coupling constant. The diagonal element of the coupling matrix are $g_{ii}= - \sum_{j\neq i} g_{ij}$. Thus, the coupling matrix satisfies the condition $\sum_{j}g_{ij}=0$ which fulfills the condition for invariance of the synchronization manifold~\cite{Pecora1998}. Let the parameter $r^i = \tilde{r}+\delta r^i$, where $\tilde{r}$ is some typical value of the parameter and $\delta r^i$ is a small mismatch.

When the coupled systems are identical, i.e. $r^i = r;\; \forall i$, the they can exhibit complete synchronization for suitable coupling constant $\varepsilon$~\cite{Pecora1990}. For complete synchronization all the state variables of the coupled systems become equal, i.e. $x^i=x;\; \forall i$ and the motion of the coupled systems are confined to the subspace defined by $x^i =x$ and this subspace is the synchronization manifold. The complementary space defines the transverse manifold. The synchronized state is stable when all the transverse Lyapunov exponents are negative. The Lyapunov exponents are calculated by expanding around the synchronous solution $x^i = x$.

For coupled nonidentical systems, the synchronization is of the generalized type, where the state variables of the coupled systems are related by a functional relationship~\cite{Abarbanel1995}. Here, we expand Eq.~(\ref{N-systems}) around the solution $\tilde{x}$ of a system with some typical parameter $\tilde{r}$ \cite{typical-r}. In the expansion, we retain terms up-to second order and we get~\cite{Acharyya2012},
\begin{eqnarray}
\dot{z}^i &=& D_x f(\tilde{x},\tilde{r}) z^i + D_r f(\tilde{x},\tilde{r}) \delta r^i + \frac{1}{2} D_r^2 f(\tilde{x},\tilde{r}) (\delta r^i)^2\nonumber \\
&&  + D_r D_x f(\tilde{x},\tilde{r}) z^i \delta r^i  + \varepsilon \sum_{j=1}^N g_{ij} D_x h(\tilde{x}) z^j
\label{first-taylor-expand}
\end{eqnarray}
where $z^i = x^i - \tilde{x}$. In Eq.~(\ref{first-taylor-expand}) we have dropped the term containing $(z^i)^2$ as we are interested in the solution $z^i \rightarrow 0$. Eq.~(\ref{first-taylor-expand}) contains both inhomogeneous and homogeneous terms. In Ref.~\cite{Acharyya2012}, we had argued that the exponents in the expanding and contracting solutions are determined by the homogeneous terms and the inhomogeneous term does not contribute to these exponents. Similar observation was made in Ref.~\cite{Sorrentino2011}. So to calculate Lyapunov exponents from Eq.~(\ref{first-taylor-expand}) we drop the inhomogeneous terms to obtain
\begin{equation}
\dot{z}^i = D_x f(\tilde{x},\tilde{r}) z^i + D_r D_x f(\tilde{x},\tilde{r}) z^i \delta r^i + \varepsilon\sum_{j=1}^N g_{ij} D_x h(\tilde{x}) z^j
\label{first-homogeneous-equation}
\end{equation}
Eq.~(\ref{first-homogeneous-equation}) can be put in the matrix form as
\begin{equation}
\dot{Z} = D_x f(\tilde{x},\tilde{r}) \ Z + D_r D_x f(\tilde{x},\tilde{r}) \ Z \ R + D_x h(\tilde{x}) \ Z \ G^T
\label{first-matrix-form}
\end{equation}
where $G^T$ is the transpose of the coupling matrix $G$ and $Z=(z^1,..,z^N)$ and $R={\rm diag}(\delta r^1,...,\delta r^N)$.

Let $\gamma_j,\;j=1,..N$ be the eigenvalues of the coupling matrix $G^T$ and the corresponding left and right eigenvectors be $e_j^L$ and $e_j^R$ respectively. We multiply Eq.~(\ref{first-matrix-form}) by $e_j^R$ from right and use the $m$-dimensional vector $\phi_j=Z e_j^R$. Thus,
\begin{equation}
\dot{\phi}_j = [D_x f + \varepsilon \gamma_j D_x h] \phi_j + D_r D_x f \ Z \ R e_j^R.
\label{second-matrix-form}
\end{equation}
Eq.~(\ref{second-matrix-form}), is not in the diagonal form since in general $e_j^R$ are not eigenvalues of $R$. To circumvent this problem, we use first order perturbation theory and obtain the first order correction due to the NDP as $\nu_j = e_j^L R e_j^R$. Thus, we can approximate Eq.~(\ref{second-matrix-form}) as
\begin{equation}
\dot{\phi}_j = [D_x f + \varepsilon \gamma_j D_x h] \phi_j + \nu_j D_r D_x \phi_j.
\label{first-master-stability-equation}
\end{equation}
The above equation can be cast in the form of master stability equation by introducing two complex parameters, namely, effective coupling parameter $\alpha = \varepsilon\gamma_j$ and mismatch parameter $\nu_r=\nu_j$ \cite{change-delta-nu} as
\begin{equation}
\dot{\phi} = [D_x f + \alpha D_x h + \nu_r D_r D_x f] \phi
\label{master-stability-equation}
\end{equation}
This equation reduces to the master stability equation of Pecora and Carroll \cite{Pecora1998} for identical systems when $\nu_r=0$.

The MSF is defined as the largest Lyapunov exponent, $\lambda_{max}$, of the above master stability equation, as a function of the parameters $\alpha$ and $\nu_r$. The MSF for coupled nearly-identical systems obtained using the above formalism is an approximation to the actual values. The accuracy of MSF and the Lyapunov exponents obtained using this formalism are discussed in Ref~\cite{Acharyya2012} with numerical examples. It is found that the errors are small when the systems are synchronized.

The MSF can be used to study the stability of synchronization of any network of N coupled nearly-identical systems. For a given network, one can determine the eigenvalues $\gamma_i, i=1,\ldots,N$ of $G$ and the corresponding $\nu_i$ values. If the MSF for $\alpha= \gamma_i \varepsilon$ and $\nu_r = \nu_i$ is negative for all the transverse eigenvalues ($i=2,\ldots,N$), then the synchronization is stable.

\begin{figure}
\begin{center}
\includegraphics[width=.9\columnwidth]{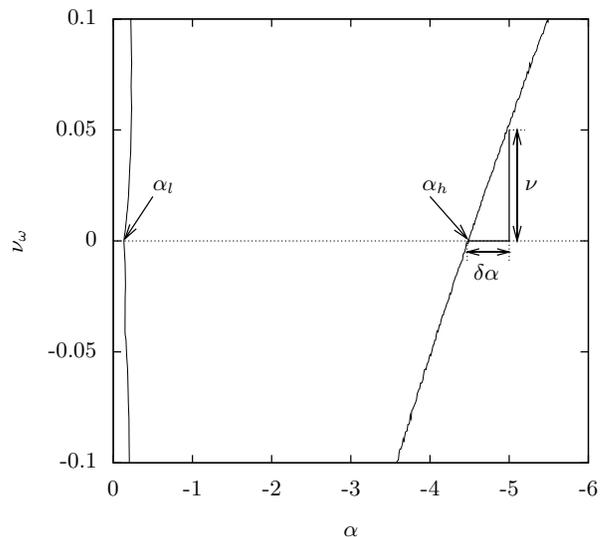}
\end{center}
\caption{\label{msfrosnonidomega} The zero contour curves of the master stability function of R\"ossler oscillators with the frequency $\omega$ as NDP are plotted in the parameter plane $\alpha-\nu_{\omega}$. In the region bounded by the zero contour curves the MSF is negative, i.e the region of stable synchronization. The range $(\alpha_l, \alpha_h)$ for $\nu_{\omega}=0$ corresponds to the range of stable synchronization for identical systems. The R\"ossler parameters are $a=b=0.2,\,c=7.0$ and $\alpha_l\sim -0.14$ and $\alpha_h\sim -4.48$. For nearly identical systems, a mismatch parameter $\nu$ corresponds to a change $\delta \alpha$ in the stability range as shown schematically in the figure. }
\end{figure}

Consider a network of coupled R\"ossler oscillators with the frequency $\omega$ as NDP and diffusive coupling in $x$ variables. The dynamics of the network is
\begin{eqnarray}
\dot{x_i} &=& -\omega_i y_i - z_i + \varepsilon \sum_j G_{ij} x_j,\nonumber\\
\dot{y_i} &=& \omega_i x_i + a y_i, \label{rossler-system} \\
\dot{z_i} &=& b + z_i(x_i-c), \nonumber
\end{eqnarray}
The master stability equations for the above system of coupled R\"ossler oscillators are
\begin{eqnarray}
\dot{\phi}_x &=& -\omega \phi_y - \phi_z + \alpha \phi_x - \nu_{\omega} \phi_y, \nonumber\\
\dot{\phi}_y &=& \omega \phi_x + a \phi_y + \nu_{\omega} \phi_x, \label{rossler-mse} \\
\dot{\phi}_z &=& z \phi_x + (x-c) \phi_z, \nonumber
\end{eqnarray}
where $\omega$ is a typical parameter. In Fig.~\ref{msfrosnonidomega} we plot the zero contours of MSF in the parameter plane $\alpha-\nu_{\omega}$ \cite{note-contour}. The region bounded by the zero contour curves corresponds to the region of negative values of MSF. If all the transverse Lyapunov exponents fall in this region then the synchronization is stable.

\subsection{\label{degenerate_cases} Degenerate eigenvalues of coupling matrix $G$}

In this section we will consider the case where the coupling matrix $G$ has degenerate eigenvalues. Degenerate eigenvalues of the coupling matrix are observed in many networks with some symmetry property, such as a star network.

When the eigenvalues are degenerate, the first order correction alone is not sufficient since the second order correction diverges. In this case one needs to use the degenerate perturbation theory.
Let the $j$-th eigenvalue $\gamma_j$ of $G^T$ have $p$ degeneracy and the left and right eigenvectors of $G^T$ corresponding to eigenvalue $\gamma_j$ be denoted by $e_{j1}^L,e_{j2}^L,...,e_{jp}^L$ and $e_{j1}^R,e_{j2}^R,...,e_{jp}^R$ respectively. For these $p$ degenerate eigenvalues, we introduce the $p\times p$ matrix $A_j$ as
\begin{displaymath}
A_j = \left(\begin{array}{cccc}
     \mu_{11} & \mu_{12} & \ldots & \mu_{1p}\\
     \mu_{21} & \mu_{22} & \ldots & \mu_{2p}\\
     \vdots & \vdots & \ddots & \vdots \\
     \mu_{p1} & \mu_{p2} & \ldots & \mu_{pp}
    \end{array}\right)
\end{displaymath}
where, $\mu_{kl}=e_{jk}^L R e_{jl}^R$. Now, we diagonalize matrix $A_j$ to get the diagonal matrix $B_j = \rm{diag}[\nu_{j1},\ldots,\nu_{jp}]$. Thus the linear stability equation (\ref{first-master-stability-equation}) can be written as
\begin{equation}
\dot{\phi}_{jk} = [D_x f + \varepsilon \gamma_j D_x h] \phi_{jk} + \nu_{jk} D_r D_x \phi_{jk};\; k=1,\ldots,p.
\label{master_stability_equation_degenerate}
\end{equation}
We note the master stability equation (\ref{master-stability-equation}) has the same form as Eq.~(\ref{first-master-stability-equation}). Thus the master stability function for the degenerate case is the same as in Sec. \ref{MSF}.

\begin{figure}
\begin{center}
\includegraphics[scale=.5]{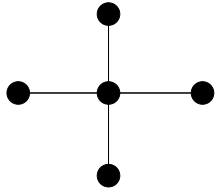}
\end{center}
\caption{\label{star_5nodes} A simple star network with 5 nodesa.}
\end{figure}

\begin{figure}
\begin{center}
\includegraphics[width=.9\columnwidth]{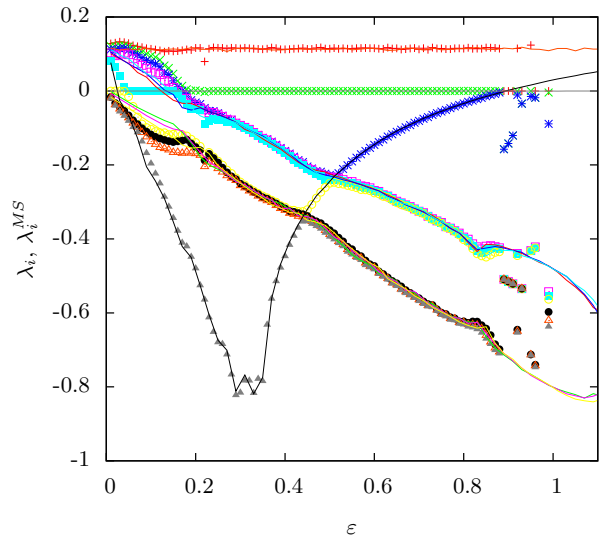}
\end{center}
\caption{\label{ale_msf_omega_star_5nodes} The $9$ largest Lyapunov exponents (points) and their estimated value (lines) using the master stability equation~(\ref{master-stability-equation}) are plotted as a function of the coupling constant $\varepsilon$ for a 5 node star network of coupled R\"ossler systems with $1\%$ mismatch in NDP $\omega$. We note that the master stability function can be used to obtain 5 Lyapunov exponents which correspond to the largest Lyapunov exponents of the master stability equation. The other Lyapunov exponents are estimated using the other Lyapunov exponents of the master stability equation.}
\end{figure}

As an example of the degenerate case we consider the star network of 5 nodes shown in Fig.~\ref{star_5nodes} . The coupling matrix $G$ is given by
\begin{eqnarray}
G_s = \left( \begin{array}{rrrrr}
-4 & 1 & 1 & 1 & 1 \\
1 & -1 & 0 & 0 & 0 \\
1 & 0 & -1 & 0 & 0 \\
1 & 0 & 0 & -1 & 0 \\
1 & 0 & 0 & 0 & -1
\end{array} \right)
\end{eqnarray}
The eigenvalues of G are, $0,-1,-1,-1,-5$. the eigenvalue $-1$ has three degeneracy.
In Fig~\ref{ale_msf_omega_star_5nodes}, nine largest Lyapunov exponents (points) and their estimated value (lines) using the master stability equation (Eq.~(\ref{master-stability-equation}) are plotted as a function of the coupling parameter $\varepsilon$ for a star network of coupled R\"ossler systems with $1\%$ variation in NDP $\omega$. We see a good agreement between the numerical and theoretical values of Lyapunov exponents in the synchronization region.

\subsection{\label{msf-numerical-results} MSF for coupled nearly-identical oscillators with more than one NDP}

\begin{figure}
\begin{center}
\includegraphics[width=.9\columnwidth]{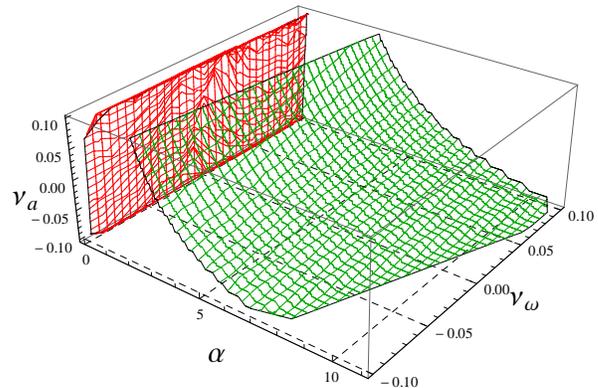}
\end{center}
\caption{\label{msf_a_omega}The zero contour surfaces (red and green) of the MSF are shown in the phase space $(\alpha,\nu_{\omega},\nu_a)$ for R\"ossler system. The MSF is negative in the region covered by these surfaces and thus it gives the stable region.}
\end{figure}

In Sec.~\ref{MSF}, we have derived the master stability equation for coupled nearly identical dynamical systems with one NDP. In this section we consider coupled nearly identical dynamical systems with more than one NDPs and derive the master stability equation for the same.

The dynamics of $i$-th oscillator can be written as
\begin{equation}
\dot{x}^i = f(x^i,r_{i1},\ldots,r_{iq}) + \varepsilon \sum_{j=1}^N g_{ij} h(x^j);\; i=1,\ldots,N
\label{dyn-net-ndp}
\end{equation}
where, $r_{i1},\ldots,r_{iq}$ are the $q$ independent NDPs of the $i$-th node. Let the typical values of these NDP's be $\tilde{r} = (\tilde{r}_1,\ldots,\tilde{r}_q)$. The linearized equation is obtained following the procedure of Sec.~\ref{MSF} and we get (see Eq.~(\ref{first-homogeneous-equation})
\begin{eqnarray}
\dot{z}^i
&=& D_x f z^i + \varepsilon \sum_{j=1}^N g_{ij} D_x h z^j + \sum_{k=1}^q D_{r_k} D_x f z^i \delta r_{ik}
\label{lin-eq-ndp}
\end{eqnarray}
where $z^i = x^i - \tilde{x}$ and $\delta r_{ik} = r_{ik} - \tilde{r}_k$.

Consider the $j$-th eigenvalue, $\gamma_j$, of the coupling matrix $G$. For $\gamma_j$, the mismatch parameter for the NDP $r_k$ using the first order perturbation correction, is $\nu_{jk} = e_j^L R_k e_j^R$ and hence Eq.~(\ref{first-master-stability-equation}) can be written as
\begin{equation}
\dot{\phi}_j = [D_x f + \varepsilon \gamma_j D_x h] \phi_j + \sum_{k=1}^q \nu_{jk} D_{r_k} D_x \phi_j.
\label{lin-eqn-corr-ndp}
\end{equation}

We can write the master stability equation as before by introducing the  effective coupling parameter $\alpha = \varepsilon \gamma_j$ and the mismatch parameters $\nu_k = \nu_{jk}, \; k=0,\ldots,q$
\begin{equation}
\dot{\phi} = [D_x f + \alpha D_x h + \sum_{k=1}^q\nu_k D_{r_k} D_x f] \phi
\label{mse-multi-ndp}
\end{equation}

As an example we consider coupled nearly identical R\"ossler oscillators with $\omega$ and $a$ as NDPs. In Fig.~\ref{msf_a_omega} The zero contour surfaces of the MSF are plotted in the three dimensional parameter space $(\alpha,\nu_{\omega},\nu_a)$, where $\nu_{\omega}$ and $\nu_a$ are the mismatch parameters corresponding to the NDPs $\omega$ and $a$ respectively. The MSF is negative in the region covered by these surfaces and thus it gives the stable region. From the figure we can see that the stable region for synchronization increases with mismatch parameter $\nu_{\omega}$ while at the same time it decreases with mismatch parameter $\nu_a$.

\section{\label{size-instability}Size-instability}

In this section we discuss the effect of NDP on size-instability of a network. As discussed in the introduction, by size instability one refers to a critical number of oscillators that can be coupled in a well structured network to obtain synchronization and  beyond this critical number no stable synchronization can be seen. A simple explanation of size-instability of coupled identical systems can be obtained from Eq.~\ref{master-stability-equation} with $\nu_r=0$. The MSF is negative in the range $(\alpha_{l},\alpha_{h})$ (see Fig~\ref{msfrosnonidomega}). For a given network we can find the stable interval of coupling parameter as $l_{\varepsilon}(N) = |\alpha_{h}/\gamma_N - \alpha_{l}/\gamma_2|$, where $\gamma_N$ and $\gamma_2$ are the minimum and maximum non-zero eigenvalues of the coupling matrix $G$ respectively.

For a given network, the synchronization is stable when the eigenvalues satisfies the condition,$\alpha_{h}/\alpha_l > \gamma_N/\gamma_2$ i.e. $l_{\varepsilon}(N) > 0$. Thus, the critical number of oscillators upto which synchronization is possible is given by
\begin{equation}
l_{\varepsilon}(N_c) = 0
\label{lnc}
\end{equation}

From the plot of MSF in Fig.~\ref{msfrosnonidomega}, we see that the NDP can play a crucial role in the size-instability since the stability range $l_{\varepsilon}(N)$ changes with the mismatch parameter $\nu$. We have obtained the mismatch parameter using the first order perturbation correction which requires the eigenvectors of $G$. Thus, the nature of the eigenvectors of $G$ plays a crucial role in determining the effect of NDP on size-instability. We will see that this leads to different effects of NDP on size-instability in star and ring networks.

\subsection{Star network}

\begin{figure}
\begin{center}
\includegraphics[width=.9\columnwidth]{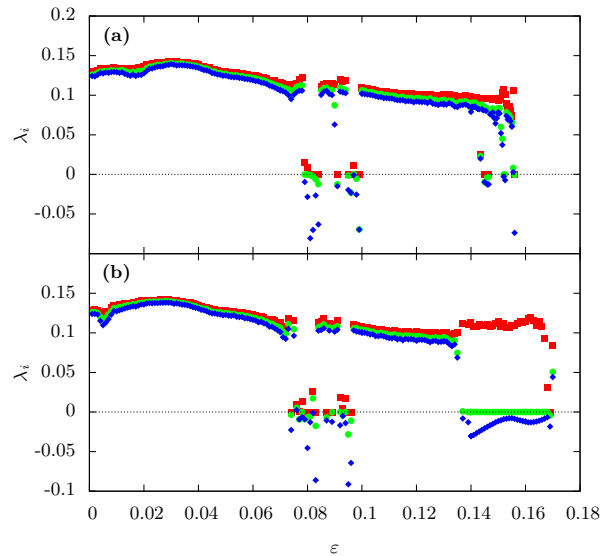}
\end{center}
\caption{\label{le32nodes} (a) Three largest Lyapunov exponents are shown as a function of $\varepsilon$ for 32 coupled identical R\"ossler oscillators on a star network. There are some windows of periodic motion between $\varepsilon = 0.075$ to $0.1$. No range of $\varepsilon$ showing stable synchronization could be detected. (b) Three largest Lyapunov exponents are shown as a function of $\varepsilon$ for 32 coupled nearly identical R\"ossler oscillators on a star network. The central node has the largest frequency $\omega = 1.05$, and the other nodes have smaller frequencies such that the average frequency of the network is $\bar{\omega} = 1.0$. A definite range of stable synchronization $\varepsilon \in (.139,.169)$ where the third largest Lyapunov exponent becomes negative, can be seen.}
\end{figure}

For a star network of $N$ nodes where all the nodes on the peripherals are only connected with a central nodes, the coupling matrix $G$ is given by
\begin{equation}
G_s = \begin{pmatrix}	
            1-N & 1 & 1 & \cdots & 1 \\
            1 & -1 & 0 & \cdots & 0 \\
            1 & 0 & -1 &  & 0 \\
            \vdots & \vdots &  & \ddots &  \\
            1 & 0 & 0 &  & -1
           \end{pmatrix},
\end{equation}
and the eigenvalues of the coupling matrix are $\gamma_1=0,\gamma_2=\cdots=\gamma_{N-1} = -1,\gamma_N = -N$. Thus, the critical number of identical oscillators that can be coupled in a star network to achieve synchronization is (Eq.~(\ref{lnc}))
\begin{equation}
N_c = \frac{\alpha_h}{\alpha_l}
\label{nc}
\end{equation}
From Fig.~\ref{msfrosnonidomega}, we get $N_c \sim 32$ for coupled identical R\"ossler systems.

Now, we consider the coupled R\"ossler oscillators with NDP $\omega$. From Fig~\ref{msfrosnonidomega}, we can see that the location of $\alpha_h$ changes as a function of the mismatch parameter $\nu$, while the location of $\alpha_l$ remains almost the same. For a star network, the eigenvector $e_N$ corresponding to the eigenvalue $\gamma_N$ is $e_N = \sqrt{\frac{1}{N(N-1)}}(N-1,-1,\ldots,-1)^T$. Thus $\nu_N = e_N^T R_{\omega} e_N = \frac{N-2}{N-1}\delta \omega_1$, where $\delta \omega_1$ is the frequency mismatch of the central node. From Fig.~\ref{msfrosnonidomega} we see that the zero contour curve near $\alpha_h$ is almost a straight line and let $b$ be the slope of this straight line. Hence,
$\alpha_h' = \alpha_h + \delta\alpha = \alpha_h + \nu_N / b$.

Thus, using Eq.~\ref{nc}, we obtain the maximum number of oscillators that can be coupled in a star network for stable synchronization for nearly identical systems as
\begin{eqnarray}
N_c' & = & \frac{\alpha_h'}{\alpha_l} \nonumber \\
& = & N_c\left[ 1 + \frac{N_c'-2}{\alpha_h b (N_c'-1)} \delta \omega_1\right]
\label{star-max-size}
\end{eqnarray}
Approximating $N_c'$ as $N_c$ in the RHS,  and using $\delta \omega_1 = 0.05$, we get $N_c' \sim 35.4$. Thus, the maximum number of oscillators that can synchronize for coupled nearly identical R\"ossler systems with $\delta \omega_1 = 0.5$ is 35.

\begin{figure}
\begin{center}
\includegraphics[width=.9\columnwidth]{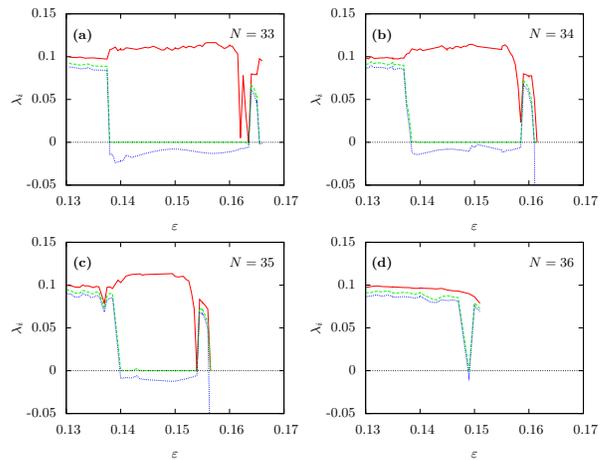}
\end{center}
\caption{\label{le_33_34_35_36nodes} (a) The three largest Lyapunov exponents are shown as a function of $\varepsilon$ for 33 coupled nearly identical R\"ossler oscillators on a star network. A finite range of $\varepsilon$ of stable synchronization is seen where the third largest Lyapunov exponent becomes negative. (b), (c), and (d) are similar figures for $N=34$, 35 and 36. We can see a finite range of $\varepsilon$ of stable synchronization for $N=34$ and~35, but not for $N=36$. In all the figures the central node has $\omega = 1.05$ and the other frequencies are chosen such that the average $\bar{\omega} = 1.0$.}
\end{figure}

We have verified the above result by explicit numerical calculations. Figs.~\ref{le32nodes}a and~\ref{le32nodes}b show the three largest Lyapunnov exponents as a function of $\varepsilon$ for 32 coupled identical R\"ossler oscillators and 32 coupled nearly identical R\"ssler oscillators respectively for a star network. For identical systems we do not see any finite range of coupling constant $\varepsilon$ showing stable synchronization while for nearly identical systems there is a finite range of $\varepsilon$  showing stable synchronization where the third largest Lyapunov exponent becomes negative. This stable range of $\varepsilon$ decreases as we increase $N$. Fig.~\ref{le_33_34_35_36nodes}a, \ref{le_33_34_35_36nodes}b, \ref{le_33_34_35_36nodes}c and \ref{le_33_34_35_36nodes}d show three largest Lyapunov exponents of coupled nearly identical R\"ssler oscillators for $N=33$, 34, 35 and 36 respectively. Thus, we see that synchronization is possible upto $N=35$ oscillators and $N=36$ does not show any 
stable synchronization.

From Eq.~(\ref{star-max-size}) we see that if we choose the frequency of the central node smaller than the average, i.e. $\delta \omega_1 < 0$ then the critical number of oscillators for stable synchronization will decrease. E.g. if $\delta \omega_1 = -0.05$, we get $N_c' = 28.6$. We have verified this result numerically.

\subsection{Ring network}
For a ring network, with nearest neighbor coupling, the coupling matrix is
\begin{equation}
G_r = \begin{pmatrix}	
            -2 & 1 & 0 & \cdots & 1 \\
            1 & -2 & 1 & \cdots & 0 \\
            0 & 1 & -2 &  & 0 \\
            \vdots & \vdots &  & \ddots &  \\
            1 & 0 & 0 &  & -2
           \end{pmatrix},
\end{equation}
The eigenvalues of $G_r$ are $\gamma_k = - 4 \sin^2\theta_k, \; \theta_k = \pi k /N, \; k=1,\ldots,N$. The corresponding eigenvectors are $e_k = \sqrt{\frac{1}{N}}(1, \exp(i2\theta_k),\ldots,\exp(i2\theta_k (N-1))^T$.

For an eigenvalue $\gamma_k$, the mismatch term due to NDP is $\nu_k = e_k^{\dagger} R e_k = \sum_j \delta r_j =0$ where we choose the average parameter as the typical value. Thus, the mismatch term is zero for all the eigenvalues $\gamma_k$. Hence, for ring network, the NDP does not have any significant effect on the size instability. We have verified this result numerically.

As noted earlier the different effect of NDP is the star and ring networks is because of the nature of eigenvectors of $G$. In the star network this can lead to either increase or decrease of the critical number of nodes for synchronization; while in the ring network it annuls the effect of NDP at least to first order and hence does not have significant effect on the critical number of nodes. 

\section{Conclusion}

In this paper we have studied the stability of synchronization of coupled nearly identical systems on a network using MSF. We extend the study to the case of degenerate eigenvalues of the coupling matrix $G$ and to more than one NDP. The main result of the paper is about the effect of NDP on size instability. The nature of the eigenfunctions of the coupling matrix $G$ play a crucial role in deciding the effect of NDP on size instability. For coupled nearly identical R\"ossler systems on a star network, we show that the critical number of nodes beyond which synchronization is not possible can be increased by having a larger frequency for the central node. For a ring network, the NDP does not have any significant effect on the critical number of nodes. In part II \cite{next} of this study we will construct optimized networks for better synchronizability and study their properties.

\section{\label{acknowledgements}Acknowledgements}

All the numerical calculations are done on the high performance computing clusters at PRL.

\end{document}